\documentclass{appolb}
\usepackage{graphicx}

\begin{document}
\title{Study of N = 16 shell closure within RMF+BCS approach
\thanks{Presented at the Zakopane Conference on Nuclear Physics “Extremes of the Nuclear
Landscape”, Zakopane, Poland, August 28 – September 4, 2016}%
}
\author{G. Saxena
\address{Department of Physics, Govt. Women Engineering College, Ajmer-305002, India}\\
M. Kaushik
\address{Department of Physics, Shankara Institute of
Technology, Jaipur-302028, India}
}
\maketitle
\begin{abstract}
We have employed RMF+BCS (relativistic mean-field plus BCS) approach to study behaviour of N = 16 shell closure with the help of ground state properties of even-even nuclei. Our present investigations include single particle energies, deformations, separation energies as well as pairing energies etc. As per recent experiments showing neutron magicity at N = 16 for O isotopes, our results indicate a strong shell closure at N = 16 in $^{22}$C and $^{24}$O. A large gap is found in between neutron 2s$_{1/2}$ and 1d$_{3/2}$ states for $^{22}$C and $^{24}$O. These results are also supported by a sharp increase in two neutron shell gap, zero pairing energy contribution and with excellent agreement with available experimental data. Moreover, our calculations of N = 16 isotones are however found at variance for higher Z isotones like $^{36}$Ca, where experiments show high lying first excited 2$^+$ state indicating shell closure at N = 16.

\end{abstract}
\section{Introduction}
Emergence of new shell closures and disappearance of conventional shell closures throughout the periodic chart
have opened various theoretical and experimental treatments in understanding the behaviour of nuclei with neutron-to-proton ratio. It has also been established that shell structure influences the locations of the neutron and proton drip
lines and the stability of matter. Appearance of new magic numbers N = 16 in the
$^{24}$O \cite{robert,kanungo} and the emergence of an N = 32
sub-shell closure in $^{52}$Ca \cite{wien} are some of the examples of changes in shell
structure. In this paper we have investigated N = 16 shell closure with the use of Relativistic Mean
Field plus BCS approach \cite{saxena,singh}.

\section{Relativistic Mean-Field Theory}
Our RMF calculations
have been carried out using the model Lagrangian density with
nonlinear terms both for the ${\sigma}$ and ${\omega}$
mesons \cite{singh}.\\
\begin{eqnarray}
       {\cal L}& = &{\bar\psi} [\imath \gamma^{\mu}\partial_{\mu}
                  - M]\psi + \frac{1}{2}\, \partial_{\mu}\sigma\partial^{\mu}\sigma
                - \frac{1}{2}m_{\sigma}^{2}\sigma^2\nonumber\\
                  && -\frac{1}{3}g_{2}\sigma
                  ^{3} - \frac{1}{4}g_{3}\sigma^{4} -g_{\sigma}
                 {\bar\psi}  \sigma  \psi -\frac{1}{4}H_{\mu \nu}H^{\mu \nu}\nonumber\\
                && + \frac{1}{2}m_{\omega}
                   ^{2}\omega_{\mu}\omega^{\mu} + \frac{1}{4} c_{3}
                  (\omega_{\mu} \omega^{\mu})^{2}
                   - g_{\omega}{\bar\psi} \gamma^{\mu}\psi
                  \omega_{\mu}\nonumber\\
               &&-\frac{1}{4}G_{\mu \nu}^{a}G^{a\mu \nu}
                  + \frac{1}{2}m_{\rho}
                   ^{2}\rho_{\mu}^{a}\rho^{a\mu}
                   - g_{\rho}{\bar\psi} \gamma_{\mu}\tau^{a}\psi
                  \rho^{\mu a}\nonumber\nonumber\\
                &&-\frac{1}{4}F_{\mu \nu}F^{\mu \nu}
                  - e{\bar\psi} \gamma_{\mu} \frac{(1-\tau_{3})}
                  {2} A^{\mu} \psi\,\,\label{eqq1}
\end{eqnarray}
where the field tensors $H$, $G$ and $F$ for the vector fields are
defined by equation (\ref{eqq1})
\begin{eqnarray}
                 H_{\mu \nu} &=& \partial_{\mu} \omega_{\nu} -
                       \partial_{\nu} \omega_{\mu}\nonumber\\
                 G_{\mu \nu}^{a} &=& \partial_{\mu} \rho_{\nu}^{a} -
                       \partial_{\nu} \rho_{\mu}^{a}
                     -2 g_{\rho}\,\epsilon^{abc} \rho_{\mu}^{b}
                    \rho_{\nu}^{c} \nonumber\\
                  F_{\mu \nu} &=& \partial_{\mu} A_{\nu} -
                       \partial_{\nu} A_{\mu}\,\,,\nonumber\
\end{eqnarray}
and other symbols have their usual meaning. Based on the
single-particle spectrum calculated by the RMF described above, we
perform a state dependent BCS calculations and continuum is replaced by a set of positive
energy states generated by enclosing the nucleus in a spherical box.
For further details of these formulations we refer the readers to
ref. \cite{singh}.
\section{Results and Discussion}
\begin{figure}[htb]
\centerline{%
\includegraphics[width=10.5cm]{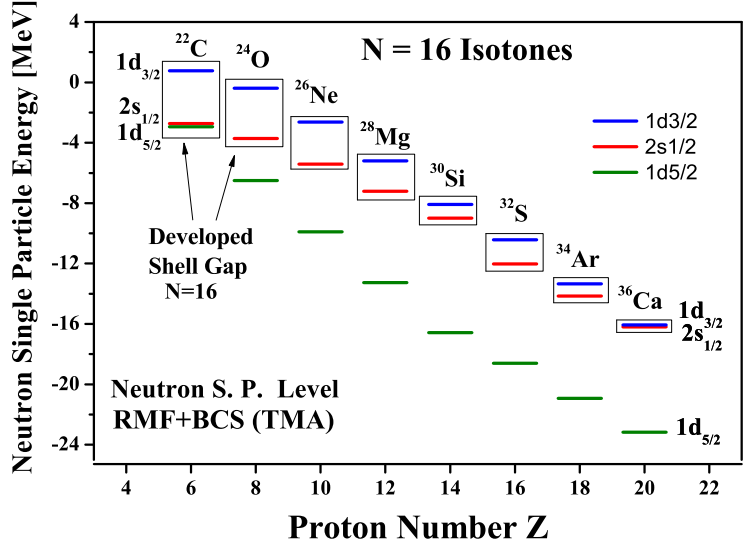}}
\caption{Single particle energies of neutron 1d$_{5/2}$, 1d$_{3/2}$ and 2s${1/2}$ states for N = 16 isotones as a function of proton number.}
\label{Fig:F2H}
\end{figure}
The results of single particle energy of N = 16 isotonic chain calculated using RMF with TMA force parameter \cite{suga} have been shown in Fig. 1.
A large variation in the energies of states 2s$_{1/2}$, 1d$_{5/2}$ and 1d$_{3/2}$ is clearly seen moving from proton rich to neutron rich (right to left).
It is evident from Fig 1 that moving towards proton deficient side 2s$_{1/2}$ state creates a substantial gap with 1d$_{3/2}$ state specially for Z = 6 and Z = 8 resulting development of new shell closure N = 16 in $^{22}$C and $^{24}$O. This gap is around 3.5 MeV and 3.3 MeV for $^{22}$C and $^{24}$O respectively as can be seen in figure. This kind of reorganization is also observed from the calculations with other parameters NL3 and PK1 (not shown here). It is gratifying to note here that our results are showing doubly magic character of $^{24}$O as observed in recent experiments \cite{robert,kanungo} and in addition the same shell closure N = 16 is also observed in $^{22}$C. On the other side, for larger Z, 2s$_{1/2}$ and 1d$_{3/2}$ states are found with very small gap giving no sign for N = 16 shell closure. This result is not in accord with experimental investigations showing shell closure at N = 16 due to high lying 2$^+$ state for $^{36}$Ca \cite{door} along with $^{30}$Si and $^{32}$S \cite{wang}. Further investigations are required for consistent description of isotonic chain in terms of parameters, pairing and isospin.
\begin{figure}[htb]
\centerline{%
\includegraphics[width=12.5cm]{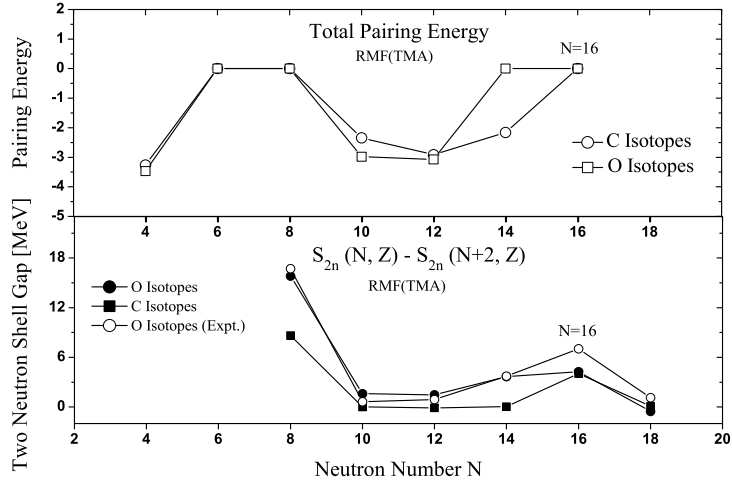}}
\caption{Lower Panel: Two Neutron Shell Gap for C and O isotopes
are compared with experimental value for O isotopes
\cite{wang}. Upper Panel: Pairing Energy for C and O isotopes.}
\label{Fig:F2H}
\end{figure}

To get into more insight, we have plotted two neutron shell gap (S$_{2n}$ (N, Z)
- S$_{2n}$ (N+2, Z)) in lower panel of Fig. 2, for C and
O isotopes calculated by RMF+BCS approach using TMA force parameter \cite{suga} along with experimental shell gap for O isotopes \cite{wang}. One can observe abrupt increase in shell gap for conventional shell
closure at N = 8. In the same way another rise in two neutron shell gap can be seen
moving from N = 14 to N = 16 for both C and O isotopes. This rise which is in accord with experiments \cite{wang} supports occurrence of new spherical shell closure at N = 16 for $^{22}$C and $^{24}$O both. Further, in upper panel of Fig. 2, we have shown paring energy contribution for both C and O isotopes.
For doubly magic nuclei pairing energy vanishes and indeed it vanishes for $^{12}$C, $^{14}$C, $^{22}$C and $^{14}$O, $^{16}$O, $^{24}$O for N = 6, 8 and 16 respectively. The result in upper panel of Fig. 2 again fortify shell closure at N = 16 for $^{22}$C and $^{24}$O and general validity of RMF approach.

\section*{Acknowledgements}
Authors are grateful to Prof. H. L. Yadav, BHU, India for his kind guidance and support. One of the authors (G. Saxena)
gratefully acknowledges the support provided by SERB (DST), Govt. of India under the young scientist project YSS/2015/000952 and International Travel Grant.

\end{document}